\documentclass[aps,prx,reprint,a4paper,floatfix,amsmath,amssymb,amsfonts,noshowpacs,longbibliography]{revtex4-2} 
\usepackage{newtxtext,newtxmath} 
\usepackage[T1]{fontenc}
\usepackage[utf8]{inputenx} 
\usepackage{graphicx} 
\usepackage[dvipsnames]{xcolor} 
\usepackage{soul}  
\usepackage[textwidth=17.5cm,textheight=23.5cm,verbose,pdftex]{geometry} 
\usepackage[pdftex]{hyperref}
\hypersetup{pdftitle={X-ray diffraction from dislocation half-loops in epitaxial films},
 pdfauthor={Vladimir Kaganer},
 bookmarksnumbered=true,colorlinks,citecolor=blue,linkcolor=blue,urlcolor=blue}

\begin{document}
\title{X-ray diffraction from dislocation half-loops in epitaxial films}
\author{Vladimir M. Kaganer}
\affiliation{Paul-Drude-Institut f\"ur Festk\"orperelektronik, Hausvogteiplatz 5--7,
10117 Berlin, Germany}

\begin{abstract}
X-ray diffraction from dislocation half-loops consisting of a misfit
segment and two threading arms extending from it to the surface is
calculated by the Monte Carlo method. The diffraction profiles and reciprocal
space maps are
controlled by the ratio of the total lengths of the misfit and the
threading segments of the half-loops. A continuous transformation
from the diffraction characteristic of misfit dislocations to that 
of threading dislocations with increasing thickness of an epitaxial film
is studied. Diffraction from dislocations
with edge and screw threading arms is considered and the contributions
of both types of dislocations are compared.
\end{abstract}
\maketitle

\section{Introduction}

Misfit dislocations are the most common mode of strain relaxation
in epitaxial films \citep{fitzgerald91,hull92,jain97,bolkhovityanov01}.
Since the dislocation lines cannot terminate inside a crystal, a misfit
dislocation is accompanied by threading arms that extend to the surface
(or terminate at an incoherent boundary; we do not consider this case
here). The glide of the threading arm under the action of epitaxial
strain is the most prominent mechanism of strain relaxation \citep{matthews74}.
Threading dislocations passing through the active region of a heteroepitaxial
structure lead to a degradation of its electronic properties, whereas
misfit dislocations, if located at the interface of a buffer layer
below the active region, may have no negative effect. Therefore,
a separate determination of misfit and threading dislocations is of
primary interest in the characterization of heterostructures for electronic
and optoelectronic applications. The density of threading dislocations
can be very low if the dislocations glide over long distances, up
to the entire length of the sample. At the other extreme, epitaxial
gallium nitride is a well-known example of a crystal with high threading
dislocation densities \citep{bennett10}.

Shifts in the positions of the X-ray diffraction peaks due to relaxation of the
average strain by misfit dislocations are commonly used to
detect strain relaxation and the corresponding misfit dislocation
density \citep{heinke94}. Dislocations also cause inhomogeneous strain,
leading to additional diffuse scattering at low dislocation densities
and to a broadening of the X-ray peaks at high dislocation densities.
The interpretation of the diffraction peak profiles is not as straightforward
as that of the mean strain due to dislocations, since the positions
of the dislocations may be correlated for kinetic or energetic reasons.
The elastic energy of a dislocation array is reduced when misfit dislocations
reduce fluctuations in the mean distances between dislocations, from
a random to a more periodic arrangement. Threading dislocations reduce
the elastic energy when dislocations with opposite Burgers vectors
are closer together to compensate for long-range strain.

The theory of X-ray diffraction from misfit \citep{kaganer97} and
threading \citep{kaganer05GaN} dislocations takes these correlations
into account and shows that the diffraction peak profiles are sensitive
to them. Scattering from misfit dislocations cannot be neglected even
in situations where the threading dislocations dominate. Reciprocal
space maps of GaN films several microns thick, where threading dislocations
are expected to dominate, also showed a significant scattering from
misfit dislocations \citep{kopp13jac,kopp14jap}. In these studies,
misfit and threading dislocations were considered as two separate
dislocation arrays uncorrelated with each other.

It is more appropriate to model the dislocation distribution by dislocation
half-loops consisting of a misfit segment and two threading arms extending
from it to the surface. The two threading segments have opposite displacement
fields, corresponding to opposite directions of the dislocation lines
when the Burgers vector is kept constant along the half-loop. Equivalently,
the two threading segments can be considered to have opposite Burgers
vectors, if the dislocation line directions are taken to be the same.
These threading dislocations screen the strain field from each other
and provide a model of the dislocation correlations that reduce the
elastic energy of the film \citep{kaganer10acta}. By varying the
relative lengths of the misfit and threading segments, one can go
from the limiting case of misfit dislocations to the opposite limit
of threading dislocations. The elastic field of a
dislocation half-loop is quite complicated (see Supporting Information)
and the diffraction from the half-loops can hardly be studied analytically.
However, the X-ray diffraction from a statistical distribution of
defects with known elastic fields can be calculated by the Monte Carlo
method \citep{kaganer09prbMC,kaganer10acta}.

The aim of the present work is to model the X-ray diffraction from
dislocation half-loops. We follow a transformation of the reciprocal
space maps and the diffraction profiles with increasing film thickness
while keeping the misfit dislocation density constant. In this way
a change from the diffraction pattern characteristic of misfit dislocations
to that of threading dislocations can be analyzed. We show that the
parameter controlling this transformation is the ratio of the total
lengths of misfit and threading dislocations, or equivalently, the
ratio of the mean length of the misfit segment to the film thickness.
We find that this transformation is rather smooth and also depends
on the inclination of the actual diffraction vector to the surface.
We compare the effects of the half-loops with the edge and screw dislocation
types of the threading arms, and find that they both contribute to
the symmetric Bragg reflections.

\section{Monte Carlo simulation of X-ray diffraction}

\begin{figure}
\includegraphics[width=1\columnwidth]{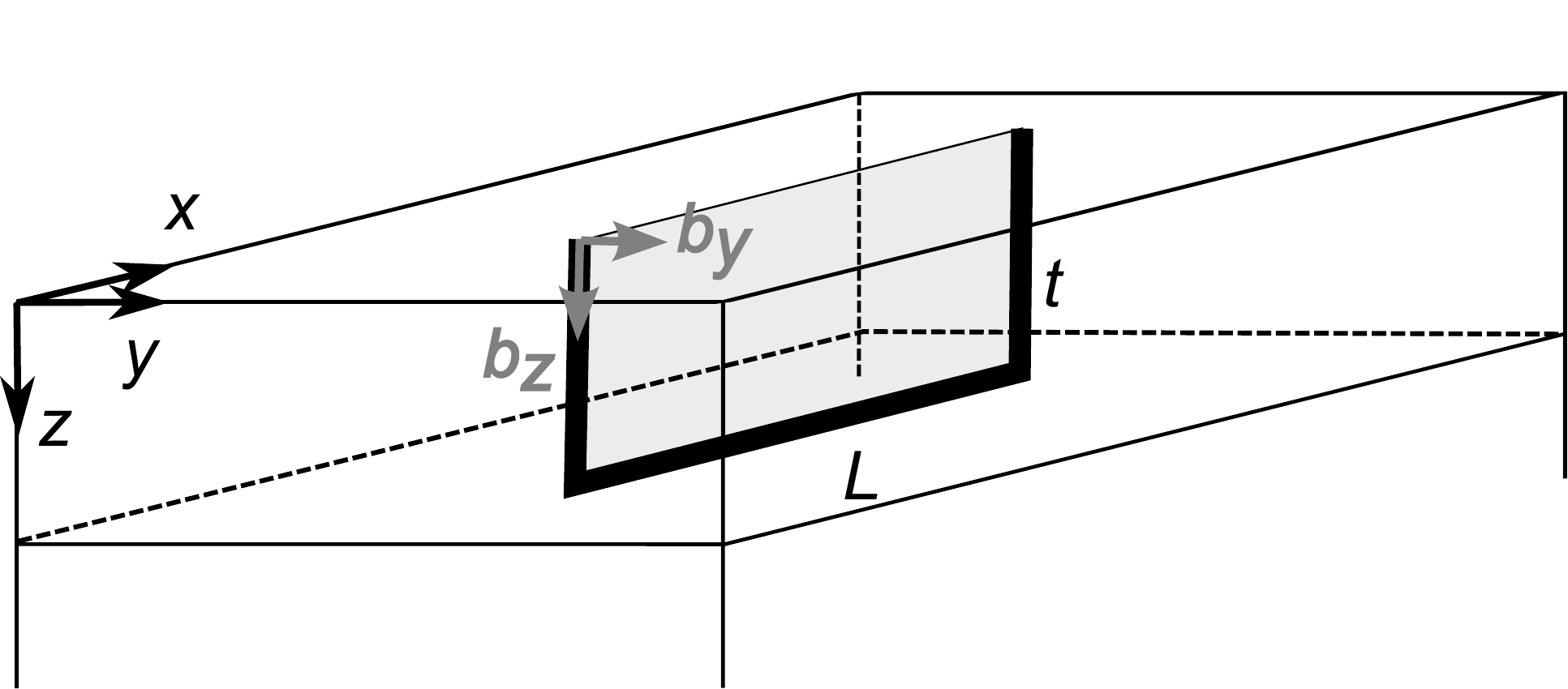}

\caption{Geometry of an epitaxial film with a dislocation half-loop.}

\label{fig:Coord} 
\end{figure}

We study the X-ray diffraction from the dislocation half-loops sketched
in Fig.\,\ref{fig:Coord}. Threading arms are assumed to be straight
and perpendicular to the film surface. Two
types of dislocations are considered. Dislocations with edge threading
arms (denoted by $b_{y}$ in Fig.\,\ref{fig:Coord}) have Burgers
vectors normal to the half-loop plane. Such half-loops correspond
to the insertion (or removal, depending on the sense of the mismatch)
of a rectangular piece of the extra atomic plane, bounded by the dislocation
line and shadowed in Fig.\,\ref{fig:Coord}. It releases the mismatch
between the film and the substrate. The second type of dislocations
has screw threading dislocation arms (denoted by $b_{z}$ in Fig.\,\ref{fig:Coord}).
Their misfit segments provide a local tilt of the film. For these half-loops,
Burgers vectors with opposite signs are taken with equal probability,
so there is no net tilt of the film.

We take the density of the threading dislocation arms $\rho_{\mathrm{T}}$ and
the mean length of the misfit segment $L$ as two parameters 
characterizing the dislocation ensemble. The misfit dislocation density
is therefore $\rho_{\mathrm{M}}=L\rho_{\mathrm{T}}/2$, since each
half-loop has two threading arms. We note that the threading dislocation
density $\rho_{\mathrm{T}}$ and the misfit dislocation density $\rho_{\mathrm{M}}$
have different dimensionalities. Threading dislocation density is
the number of threading dislocations per unit area of the surface,
or more generally, the total length of the threading dislocations
per unit volume. Misfit dislocation density is the number of dislocations
per unit length of interface, or more generally, the total length
of the dislocation lines per unit area of interface.

A parameter that controls the relative contributions of misfit and
threading dislocations is the ratio $L/t$ of the mean length of the
misfit segment $L$ to the film thickness $t$. One can also compare
the total length of misfit dislocations per unit area of the interface
$\rho_{\mathrm{M}}$ with that of threading dislocations $\rho_{\mathrm{T}}t$,
since the length of each threading segment is $t$. Given the definition
of $\rho_{\mathrm{M}}$ above, this ratio is simply $L/2t$. Another
parameter of the dislocation array is the dimensionless parameter
$M$ introduced by Wilkens \citep{wilkens70pss,wilkens76} to characterize
the screening of the dislocation strain by the surrounding dislocations.
It is equal to the ratio of the mean distance $L$ between threading
dislocations with opposite Burgers vectors (assuming the same dislocation
line directions of the threading segments) to the mean distance between
threading dislocations $\rho_{\mathrm{T}}^{-1/2}$, so that $M=L\rho_{\mathrm{T}}^{1/2}$.

The Monte Carlo simulations below are performed for an example of a GaN$\{ 0001 \}$ epitaxial film. The positions of the dislocation
half-loops are random and uncorrelated. The lengths $L$ of the misfit
segments have a lognormal distribution with the standard deviation
$L/2$. The misfit segments of the half-loops run in three equivalent
$\left\langle 1\bar{1}00\right\rangle $ directions with equal probability.
The length of the Burgers vector of a half-loop with edge threading
arms $b_{x}$ is $a=0.319$~nm, while that of the half-loop with
screw threading arms \textbf{$b_{z}$} is $c=0.518$~nm. The displacement
field of a half-loop, satisfying the elastic boundary conditions of
the free surface, is constructed from the displacement field of an
angular dislocation near the free surface \citep{Comninou1975} and
that of a dislocation normal to the surface \citep{lothe92}. Details
of the construction and the analytical expressions for all components
of the displacements are given in the Supplementary Information. 

The choice of Poisson's ratio to model dislocations in GaN is somewhat
ambiguous. In strain relaxation problems for elastically anisotropic
epitaxial films, the Poisson ratio is commonly chosen to give the
same vertical strain as in the isotropic approximation. For GaN(0001)
this requirement gives $\nu=c_{13}/(c_{13}+c_{33})$, where $c_{ij}$
are the anisotropic elastic moduli. The value $\nu=0.21$ is obtained
using elastic moduli of GaN from Ref.\,\citep{polian96}. The measured values of $\nu$
vary from 0.15 to 0.23 \citep{moram09}. On the other hand, the strain field of a
straight edge dislocation with $\left\langle 0001\right\rangle $
dislocation line direction in an anisotropic hexagonal crystal coincides
with the isotropic solution when the Poisson ratio is taken to be
$\nu_{h}=c_{12}/(c_{12}+c_{11})$ \citep{belov92}. Using the elastic
moduli of GaN \citep{polian96}, Poisson's ratio is $\nu_{h}=0.27$.
We use the latter value in the Monte Carlo simulations below, to get
a better representation of the strain fields of the threading dislocation
arms.

The diffracted intensity is a Fourier transform of the correlation
function $G(\mathbf{r}_{1},\mathbf{r}_{2})=\left\langle \exp\left[i\mathbf{Q}\cdot\left(\mathbf{U}(\mathbf{r}_{2})-\mathbf{U}(\mathbf{r}_{1})\right)\right]\right\rangle $
to reciprocal space. Here $\mathbf{r}_{1}$ and $\mathbf{r}_{2}$
are the coordinates of two points inside the crystal, $\mathbf{U}(\mathbf{r})$
is the total displacement due to all dislocations (equal to the sum
of the displacement fields of individual dislocations due to linear
elasticity) calculated in these two points, and $\mathbf{Q}$ is the
diffraction vector. The statistical average $\left\langle \ldots\right\rangle $
over the dislocation ensemble and the Fourier transform can be performed
simultaneously in one and the same Monte Carlo integration \citep{kaganer09prbMC}.
This integration is time consuming, especially when dislocation densities
are large and low intensities at asymptotes are of interest: the integration
is a summation of complex numbers of modulus 1 to finally obtain a
real number which is much less than one.

When the dislocation density is large and hence the mean-squared strain
is large, only correlations between closely spaced points $\mathbf{r}_{1}$
and $\mathbf{r}_{2}$ are of importance \citep{krivoglaz63}. The
expansion $\mathbf{Q}\cdot\left(\mathbf{U}(\mathbf{r}_{2})-\mathbf{U}(\mathbf{r}_{1})\right)\approx(\mathbf{r}_{2}-\mathbf{r}_{1})\cdot\nabla(\mathbf{Q}\cdot\mathbf{U})$
allows to reduce the X-ray intensity calculation to the calculation
of the probability density of the respective distortion components
\citep{StokesWilson44,kaganer14acta}. Specifically, the intensity
$I(q_{x},q_{z})$ in the reciprocal space map is calculated as the
joint probability density of the distortions $q_{x}=-\partial(\mathbf{Q}\cdot\mathbf{U})/\partial x$
and $q_{z}=-\partial(\mathbf{Q}\cdot\mathbf{U})/\partial z$. These
distortions depend on a depth $z$ of the point in the epitaxial film
at which they are calculated. Therefore, an integration over $z$
is performed from the surface $z=0$ to the interface $z=t$. As is
usual for a Monte Carlo simulation, this integration does not require
any additional computational effort: the point $z$ is randomly and
homogeneously seeded on the interval $[0,t]$. Similarly, the intensity
$I(q)$ in a double-crystal scan with an open detector, in particular
the scan in skew geometry, is calculated as the probability density
of the distortion $q=-\hat{\mathbf{K}}^\textrm{out}\cdot\nabla(\mathbf{Q}\cdot\mathbf{U})$,
where $\hat{\mathbf{K}}^\textrm{out}$ is a unit vector in the direction
of the diffracted beam and the integration of the probability density
over $z$ is performed, as above. This expression for the distortion
component was derived in Appendix A of Ref.\,\citep{kaganer15jpd}
and is re-derived in the Supplementary Information in more familiar
Cartesian coordinates. The wave vector $q$ is related to the angular
deviation $\omega$ by $q=Q\omega\cos\theta$, where $\theta$ is
the Bragg angle of the actual reflection. The calculation of the strain
probability density distribution is orders of magnitude faster than
the straightforward calculation mentioned above because it avoids
summing the oscillating complex terms.

This Monte Carlo calculation is ideally suited to parallel computing
as each realization of the random dislocation distribution can be
computed independently and the partial sums obtained on different
processors can be added at the end. We use the $\mathtt{coarray}$
extension to Fortran, which was added to the language standard in
2008. In practice, the parallel computations require only a few lines
of code to be modified and are performed on 128 cores without any
loss of computational efficiency.

Monte Carlo simulations are performed on an 
Epyc\texttrademark\ 7763 compute server. Diffraction profiles and maps 
are typically computed in less than 1 minute with sufficient accuracy to reveal
the features of the intensity distribution. Each of the curves and 
maps presented below took several hours to reduce the statistical noise. 
As the statistical error decreases as $1/\sqrt{N}$, where $N$ is the number of repetitions, 
the one-minute runs are only an order of magnitude less accurate in intensity. 
The computation time can be reduced by choosing larger steps in the angles
in the curves and wave vectors in the maps. On the other hand, most of
the computation time is the calculation of the dislocation displacements by analytical
formulae presented in the Supplementary Information, which leaves very little room for
improvement. The calculation requires memory for an array of the calculated intensity and 
an array of the coordinates of the dislocation in an actual realization of their distribution, which together
do not exceed several megabytes per core.

\section{Results}

\begin{figure*}
\includegraphics[width=1\textwidth]{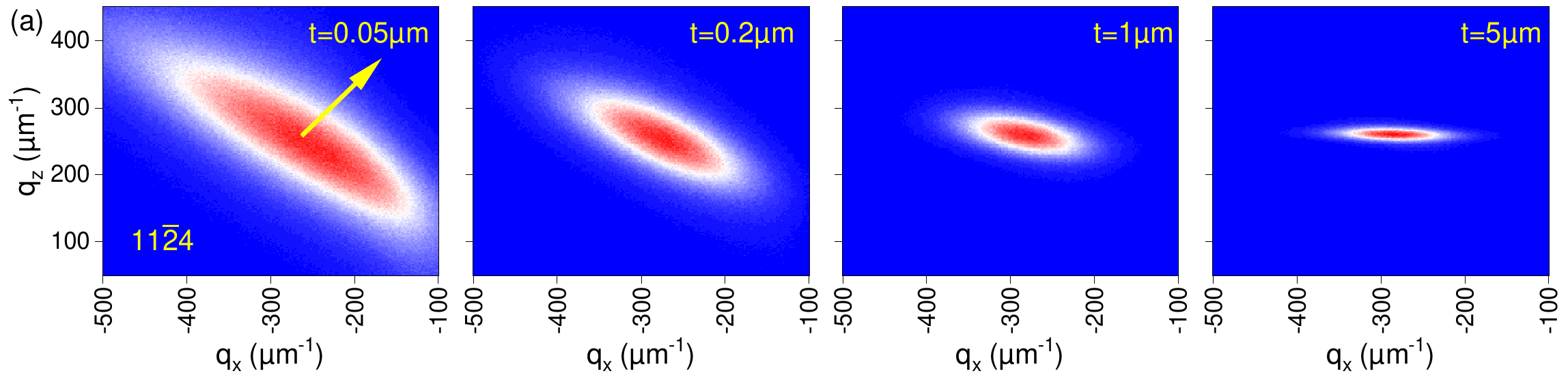}

\includegraphics[width=1\textwidth]{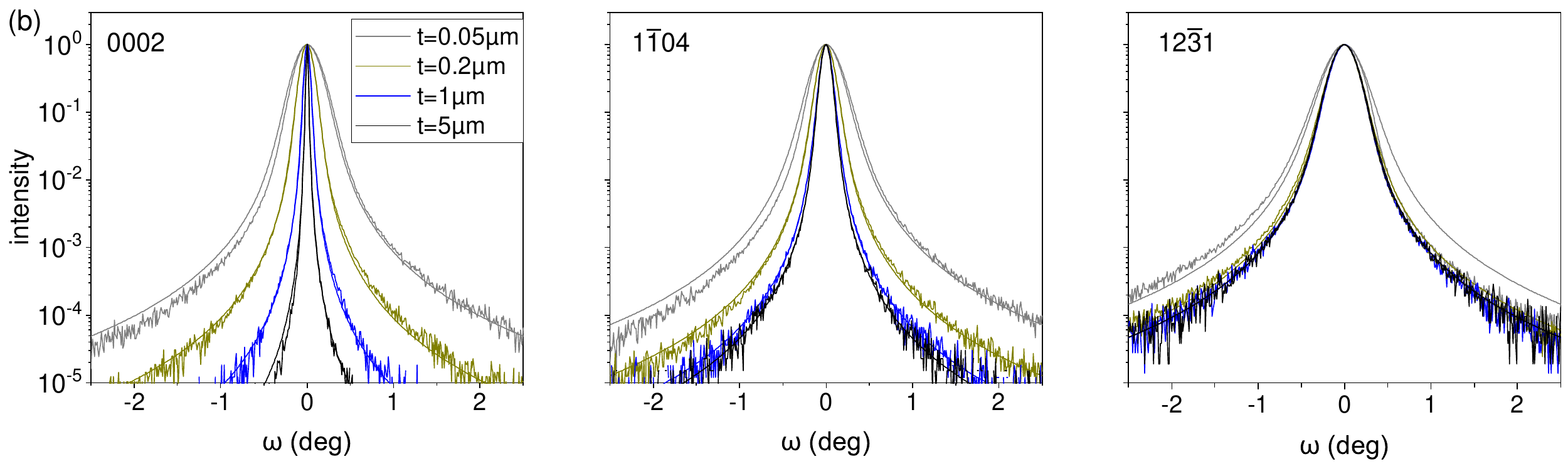}

\includegraphics[width=1\textwidth]{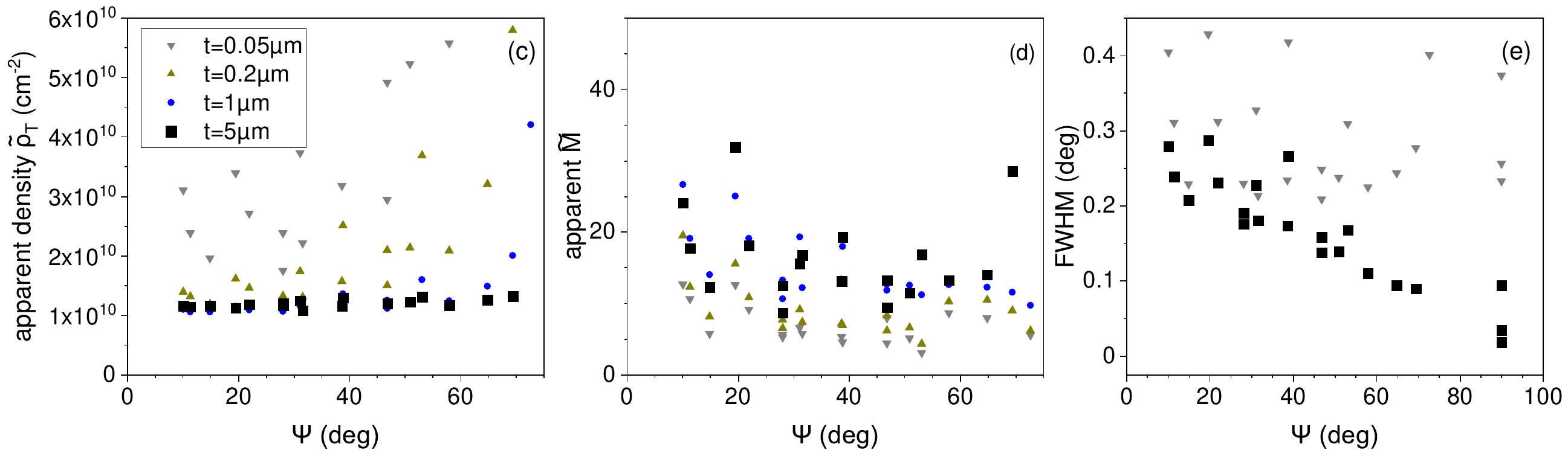}
\caption{Monte Carlo calculation of the X-ray diffraction from dislocation
half-loops with edge threading arms. Threading dislocation density
$\rho_{\mathrm{T}}=1\times10^{10}$\,cm$^{-2}$, mean length of the
misfit segments $L=1$\,\textmu m. (a) Reciprocal space maps in $11\bar{2}4$
reflection for different epitaxial layer thicknesses. The diffraction
vector is indicated by an arrow on the left map. (b) Diffraction peak
profiles in skew geometry. The noisy lines are Monte Carlo simulations,
while the smooth curves are fits that treat the diffraction intensity
as due only to threading dislocations. (c) Apparent density of threading
dislocations $\widetilde{\rho}_{\mathrm{T}}$ and (d) apparent values
$\widetilde{M}$ of the Wilkens parameter obtained in these fits.
(e) The full width at half maximum of the diffraction profiles (FWHM)
of the reflections. $\Psi$ is the angle between the reflection vector
and the crystal surface.}

\label{fig:threading} 
\end{figure*}

Let us consider the X-ray diffraction from dislocation half-loops
with edge threading arms. We assume a threading dislocation density
$\rho_{\mathrm{T}}=1\times10^{10}$\,cm$^{-2}$ and a mean length
of the misfit segments $L=1$\,\textmu m. Figure \ref{fig:threading}(a)
shows a transformation of the $11\bar{2}4$ reciprocal space maps
with increasing film thickness. For a thickness $t=0.05$\,\textmu m,
which is small compared to the misfit segment length, the misfit dislocations
dominate the diffraction. The reciprocal space map has the same features
as that of infinitely long misfit dislocations \citep{kaganer97}.
It is extended in the direction almost perpendicular to the direction
of the diffraction vector indicated by an arrow in the figure (these
directions need not be exactly perpendicular to each other since as
this is not required by symmetry). In the opposite limit, where the
thickness $t=5$\,\textmu m is large compared to the misfit segment
length, the diffraction is dominated by threading dislocation arms.
Since threading dislocations are parallel straight lines in real space,
their diffraction pattern in the reciprocal space is a disc perpendicular
to the dislocation line \citep{kopp13jac,kopp14jap}. A section of
the disc through the scattering plane gives the horizontal streak
in the map. The maps in Fig.~\ref{fig:threading}(a) show a gradual
transition from one limit to the other. At thickness $t=0.2$\,\textmu m,
five times smaller than the misfit segment length, the diffraction
pattern already differs from that for misfit dislocations. At thickness
$t=5$\,\textmu m, five times larger than the misfit segment length,
there is still a finite width of the intensity spot in the $q_{z}$
direction.

Figure \ref{fig:threading}(b) shows diffraction profiles in skew
geometry \citep{srikant97,sun02,kaganer05GaN} for the same film thicknesses
and for three reflections, a symmetric  $0002$ reflection (left),
a slightly asymmetric $1\bar{1}04$ reflection (middle), and a highly
asymmetric $12\bar{3}1$ reflection (right). The intensities calculated
by the Monte Carlo method are seen as noisy lines, while smooth lines
of the same colors are the fits discussed below. Let us start the
analysis with the symmetric reflection. Since straight edge dislocations
in an infinite medium produce strain only in the plane normal to the
dislocation line, it is expected that edge threading dislocations
do not cause any broadening of the symmetric reflections. However,
the plot in Fig.\,\ref{fig:threading}(b) shows that the total effect
of the strain field
of the misfit segment and the strain due to stress relaxation at the
free surface of the threading segments of the half-loop give rise
to a diffraction peak broadening even at a thickness of 5\,\textmu m. 

In the usual treatment of the broadening of the symmetric reflections
as a manifestation of the screw dislocations, this broadening would
be interpreted as a density of screw dislocations. The smooth lines
in the plots of Fig.\,\ref{fig:threading}(b) are the fits proposed
in Ref.\,\citep{kaganer05GaN}. They include two parameters, the dislocation
density and the length of the strain field screening (or the dimensionless
parameter $M$). An apparent density of screw threading dislocations,
obtained in the fit of the 0002 reflection for the film thickness
of 5\,\textmu m, is $1.1\times10^{8}$\,cm$^{-2}.$ The apparent
density of screw dislocations increases with the decreasing film thickness,
as can be seen from the plots, and reaches $6.5\times10^{9}$\,cm$^{-2}$
for a film thickness of 0.05\,\textmu m.

At the opposite extreme of a highly asymmetric $12\bar{3}1$ reflection
in the right plot of Fig.\,\ref{fig:threading}(b), the strain due
to edge threading arms dominates. The diffraction profiles almost
coincide for the film thicknesses of $0.2$\,\textmu m and above.
A slightly asymmetric $1\bar{1}04$ reflection in the middle plot
of Fig.\,\ref{fig:threading}(b) shows an intermediate behavior:
for thicknesses less than 1\,\textmu m, the misfit segment of the
half-loop makes a significant contribution.

Figures \ref{fig:threading}(c) and \ref{fig:threading}(d) summarize
the results of the fits made by the model for infinitely long edge
threading dislocations \citep{kaganer05GaN}. These fits are represented
by smooth lines in Fig.\,\ref{fig:threading}(b). A total of 19 diffraction
profiles in different asymmetric reflections in skew geometry
are calculated by the Monte Carlo method. The apparent density of
edge threading dislocations $\widetilde{\rho}_{\mathrm{T}}$ and the
corresponding apparent parameter $\widetilde{M}$ are obtained in
the fits. The results for different reflections are compared by plotting
these apparent parameters as a function of the angle $\Psi$ between
the diffraction vector and the film surface. $\Psi=0$ corresponds
to diffraction in the surface plane, and $\Psi=90^{\circ}$ to symmetric
reflections. The symmetric reflections are not included in Figs.\,\ref{fig:threading}(c)
and \ref{fig:threading}(d) since they have been fitted to screw rather
than edge threading dislocations.

The results for the film thickness of 5\,\textmu m are shown in Figs.\,\ref{fig:threading}(c)
and \ref{fig:threading}(d) by full squares, deliberately made larger
than the symbols for the other thicknesses, as they come closest to
the model of infinite threading dislocations assumed by the fits.
The dislocation density obtained in the fit for this film thickness
is quite close to the density of $1\times10^{10}$\,cm$^{-2}$ modeled
in the Monte Carlo simulations. This result confirms the consistency
between the present Monte Carlo simulations and the fits by the 
formulae from Ref.\,\citep{kaganer05GaN}.
Figure \ref{fig:threading}(c) shows that as the thickness decreases,
the misfit parts of the half-loops make progressively larger contributions.
The apparent density of edge dislocations can be 6 times larger than
the real density. It can also be seen that the apparent density systematically
depends on the inclination angle $\Psi$ of the reflection: the less
asymmetric reflections give a larger apparent density. This dependence
can help to recognize the contribution of misfit dislocations.

The input value of the parameter $M$ in the Monte Carlo simulations
is $M=L\rho_{\mathrm{T}}^{1/2}=10$. The values obtained in the fit
are several times larger and show a large scatter even for the 5\,\textmu m
film thickness, where the threading dislocations dominate. This result
is not surprising: as it was discussed in Ref.\,\citep{kaganer05GaN},
the fit does not take into account the orientation factors involved
in this parameter. As a result, the accuracy of the dislocation correlation
determination is lower than that of the dislocation density determination.
As it has been also discussed in Ref.\,\citep{kaganer10acta}, the
consideration of these orientation factors is a rather complicated
task. On the other hand, it is the dislocation density rather than
the dislocation correlations that is of primary interest.

Figure \ref{fig:threading}(e) shows the full widths at half maxima
(FWHMs) of the peaks obtained from the Monte Carlo simulation. The
data are shown for the film thicknesses of 0.05\,\textmu m and 5\,\textmu m.
The points from the intermediate thicknesses (not shown) are scattered
in between. The FWHMs are used to estimate the dislocation density
by a popular, because of its extreme simplicity, formula $\rho_{\mathrm{T}}=\mathrm{FWHM}^{2}/4.35b^{2}$
\citep{metzger98}. This formula is used in symmetric or asymmetric
reflections with the Burgers vector $b$ equal to either $c$ or $a$
lattice parameter of GaN to obtain the densities of either screw or
edge dislocations. The correct use of this formula for edge dislocations
implies the use of twist, i.e., extrapolation of the peak widths in
Fig.\,\ref{fig:threading}(e) to $\Psi=0$ \citep{sun02}.

When the threading dislocation arms are long and dominate in the scattering
(full squares), the FWHMs of the asymmetric reflections in Fig.~\ref{fig:threading}(e)
increase with the increasing inclination of the reflection (the angle
$\Psi$ decreases). The same dependence is observed in experiments
\citep{heinke00,sun02,kaganer05GaN}. Extrapolation to $\Psi=0$ gives
a ``twist'' of $0.3^{\circ}$, which according to the above formula
gives a threading dislocation density of $6\times10^{9}$\,cm$^{-2}$,
about half of the threading dislocation density used on input in
the Monte Carlo simulations. Thus, this simple formula gives a reasonable
estimate of the threading dislocation density, with some underestimation.
Further Monte Carlo simulations (not presented here) show that this
underestimation is systematic. The reflections for a thin epitaxial
film, shown by triangles in Fig.~\ref{fig:threading}(e), give a
large scatter of the FWHMs of different reflections and, on average,
a similar ``twist''. Hence, the FWHM based determination of the threading
dislocation density gives the same underestimate. The FWHMs of the
symmetric reflections, shown by the points at $\Psi=90^{\circ}$ in
Fig.~\ref{fig:threading}(e), depend significantly on the order of
the reflections. The 0002 reflection would give an apparent density
of screw dislocations of $1\times10^{7}$\,cm$^{-2}$ for the 5\,\textmu m
thick film and $4\times10^{9}$\,cm$^{-2}$ for the 0.05\,\textmu m
thick film. 

\begin{figure}
\includegraphics[width=1\columnwidth]{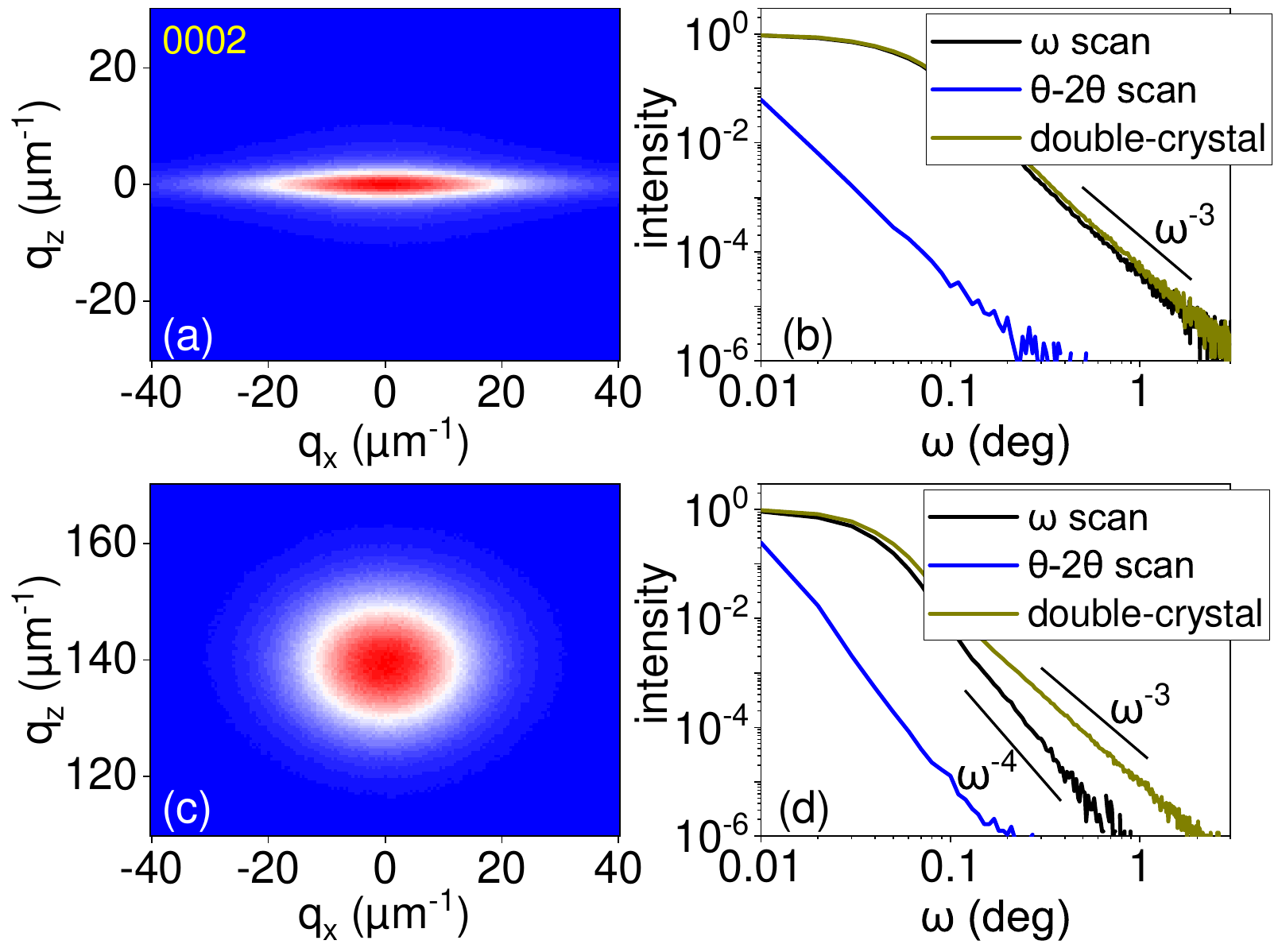}

\caption{Reciprocal space maps in a symmetric Bragg reflection $0002$ from
dislocation half-loops with (a) screw threading arms, $\rho_{\mathrm{T}}=1\times10^{9}$\,cm$^{-2}$,
and (c) edge threading arms, $\rho_{\mathrm{T}}=1\times10^{10}$\,cm$^{-2}$.
The mean length of the misfit segments is $L=1$\,\textmu m, the
film thickness is $t=1$\,\textmu m. The $\omega$ and $\theta$-2$\theta$
triple-crystal scans through the maps, as well as the double-crystal
scans, are shown in (b) and (d) in log-log scale.}

\label{fig:SymmRefl} 
\end{figure}

Figure \ref{fig:SymmRefl} shows the reciprocal space maps and the
diffraction profiles in symmetric 0002 reflection of dislocation half-loops
with screw (top) and edge (bottom) threading arms. In both cases the
mean length of the misfit segments and the film thickness are taken
to be the same, $L=1$\,\textmu m and $t=1$\,\textmu m. The dislocation
densities differ by an order of magnitude, half-loops with screw threading
arms of density $\rho_{\mathrm{T}}=1\times10^{9}$\,cm$^{-2}$ are
compared with half-loops with the edge threading arms of density $\rho_{\mathrm{T}}=1\times10^{10}$\,cm$^{-2}$.

The screw threading arms dominate in the diffraction pattern of the
respective half-loops, since the displacement due to a screw dislocation
is along the diffraction vector. As a result, the diffraction intensity
in the map of Fig.\,\ref{fig:SymmRefl}(a) is extended in the lateral
direction, perpendicular to the direction of the screw arms. The scan
in the $q_{x}$ direction in the map, which coincides with the $\omega$
scan in the symmetric reflection, collects all the diffracted intensity.
Figure \ref{fig:SymmRefl}(b) shows that the $\omega$ scan and the
double crystal scan almost coincide and have the expected $\omega^{-3}$
asymptote. 

The edge threading arms contribute to diffraction in a symmetric Bragg
reflection only due to the strain resulting from elastic relaxation
at the free surface, since the displacement field of the edge threading
dislocation in infinite medium is perpendicular to the diffraction
vector. The intensity in the reciprocal space map in Fig.\,\ref{fig:SymmRefl}(c)
is mainly due to the misfit segments of the half-loops and extends
in both $q_{x}$ and $q_{z}$ directions. The intensity in the $\omega$
scan shown in Fig.\,\ref{fig:SymmRefl}(d) has an $\omega^{-4}$
asymptote, while the additional integration in the reciprocal space
for the double-crystal scan gives rise to an $\omega^{-3}$ dependence.

Comparing the double crystal scans in Figs.\,\ref{fig:SymmRefl}(b)
and \ref{fig:SymmRefl}(d), one can see that $1\times10^{9}$\,cm$^{-2}$
half-loops with screw threading arms and $1\times10^{10}$\,cm$^{-2}$
half-loops with edge threading arms give very close diffraction curves.
Thus, when dislocation half-loops with comparable lengths of the misfit
and threading segments are present, the common assumption that the
intensity in symmetric Bragg reflections is due to screw threading
dislocations and the intensity in asymmetric reflections is due to
edge threading dislocations, is no longer valid.

\section{Conclusions}

The use of the displacement field of an angular dislocation allows
the construction of arbitrary dislocation arrangements in epitaxial
films, in particular dislocation half-loops. The X-ray diffraction
of an epitaxial film with an arbitrary density of dislocation half-loops
can be calculated by the Monte Carlo method. For
large dislocation densities and significant broadening of the diffraction
peaks, the diffraction intensity can be calculated as the probability
density of the corresponding strain components, in the Stokes-Wilson
approximation. The use of this approximation allows the calculation
time to be reduced by several orders of magnitude.

The shape of the double-crystal diffraction curves for half-loops is 
the same as that for threading dislocations. When both misfit and threading 
dislocations are present, a joint analysis of the double crystal diffraction curves
in skew geometry and reciprocal space maps in coplanar geometry is 
required to distinguish their contributions.

X-ray diffraction from dislocation half-loops is controlled by the
ratio of the total lengths of the misfit and the threading segments.
A significant deviation from the scattering pattern of misfit dislocations
is already seen in the reciprocal space maps when this ratio is 5:1,
and the opposite limit of threading dislocations is not yet reached
when this ratio is 1:5. 
An apparent density of threading dislocations obtained by fits to
the formula derived for threading dislocations alone is up to 6 times larger
than the real density of the threading segments. The apparent density
obtained in this way scatters significantly depending on the reflection
chosen. This scatter in density can be used to distinguish between
half-loops and only threading dislocations. Another indicator that
may help to distinguish between these two cases is the dependence
of the FWHMs of the reflections on the angle $\Psi$ between the reflection
vector and the surface. For threading dislocations the FWHMs increase
as $\Psi$ decreases. When misfit dislocations dominate, the FWHMs
show a larger scattering without a systematic $\Psi$ dependence.

For the half-loops with comparable total lengths of the misfit and
threading segments, both the half-loops with edge and screw threading
arms contribute to the diffraction curves in symmetric Bragg reflections.
The contribution of the half-loops with screw threading arms is an
order of magnitude larger for comparable dislocation densities. However,
since the densities of the screw threading dislocations in GaN films
grown by molecular beam epitaxy is an order of magnitude smaller than
those of edge dislocations, the contributions of both dislocation
types are comparable. In this case a clear distinction between the
dislocation types can be seen in the reciprocal space maps: the diffraction
spot for half-loops with edge threading arms is roundish, while for
those with screw arms it is laterally elongated.

\begin{acknowledgments}
The author thanks Oliver Brandt for providing access to and maintaining 
the compute server that was used for the Monte Carlo simulations in this study, 
as well as for  many useful discussions and a critical reading of the manuscript.
\end{acknowledgments}


%

\end{document}


\title{Supplementary Information to the paper \\
 X-ray diffraction from dislocation half-loops in epitaxial films}
\author{Vladimir M. Kaganer}
\affiliation{Paul-Drude-Institut f\"ur Festk\"orperelektronik, Hausvogteiplatz 5--7,
10117 Berlin, Germany}

\maketitle

\section{ Construction of the displacement field of a dislocation half-loop}

\begin{figure}
\includegraphics[width=1\columnwidth]{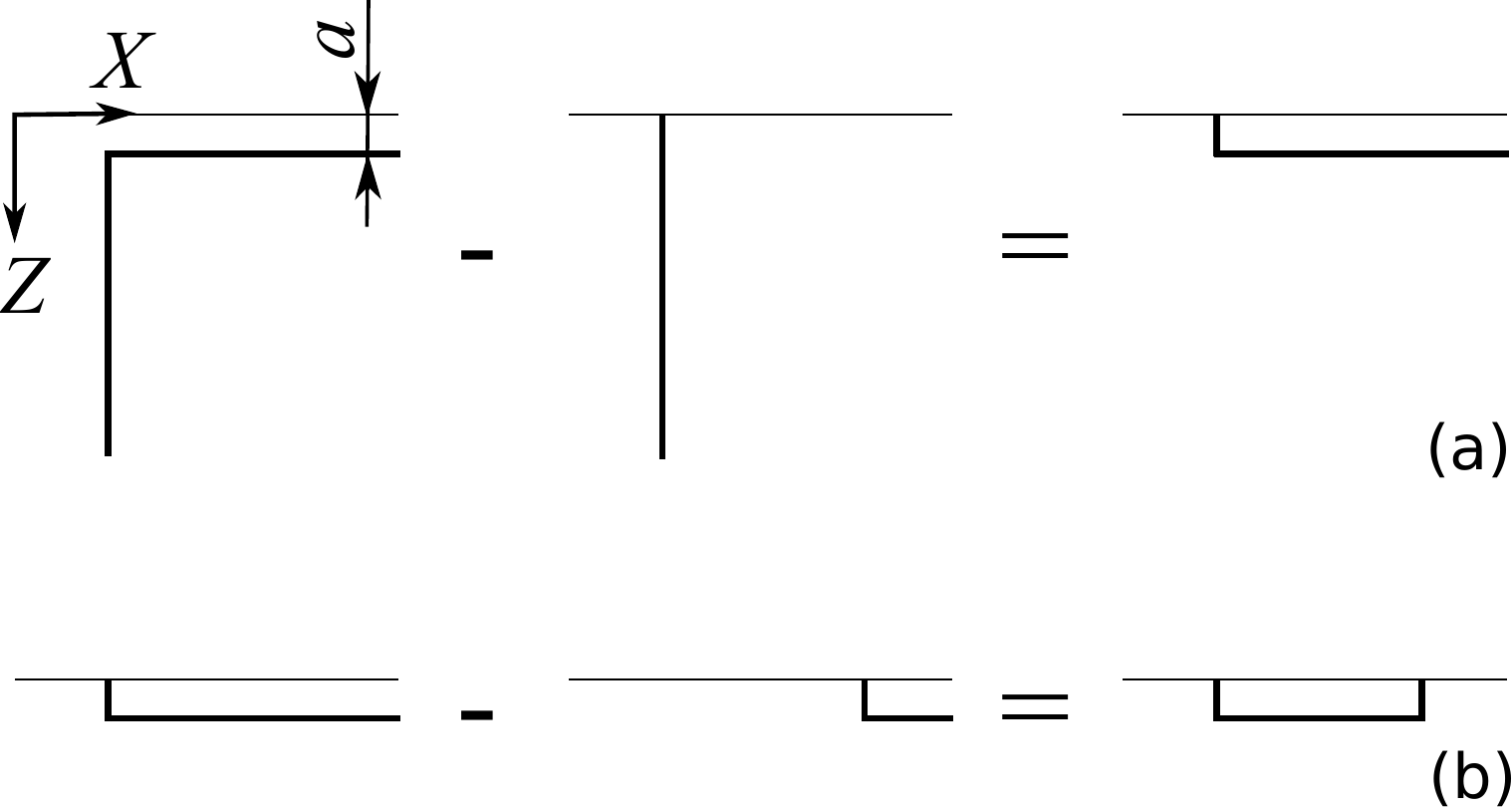}

\caption{Construction of the displacement field of a dislocation half-loop:
(a) subtract the displacement field of a dislocation perpendicular
to the surface from the displacement field of an angular dislocation
to obtain an L-shaped dislocation and (b) take the difference between
the displacement fields of two L-shaped dislocations shifted with
respect to each other. All dislocations have the same Burgers vector.}

\label{fig:ConstructHalfloop}
\end{figure}

Figure \ref{fig:ConstructHalfloop} shows a construction of the displacement
field of a dislocation half-loop. The building blocks are an angular
dislocation with one arm parallel to the surface and the other arm
perpendicular to it, and a straight dislocation perpendicular to the
surface. The explicit expressions for the respective displacement
fields are given in Sec.\,\ref{sec:AngularDisl} and \ref{sec:ThreadingDisl}
below. First, we construct an L-shaped dislocation consisting of a
half-infinite dislocation parallel to the surface and a finite dislocation
segment from it to the surface, see Fig. \ref{fig:ConstructHalfloop}(a).
It is obtained by subtracting the displacement field of the straight
dislocation from that of the angular dislocation. The difference between
the displacements of two such L-shaped dislocations, lying in the
same plane and shifted with respect to each other, gives rise to a
dislocation half-loop with the misfit dislocation segment parallel
to the surface and two threading dislocation segments from it to the
surface, as shown Fig. \ref{fig:ConstructHalfloop}(b).

\section{Displacement field of an angular dislocation with one arm parallel
to the surface  and the other arm \\ perpendicular to it}

\label{sec:AngularDisl}

Comninou and Dundurs \citep{Comninou1975} derived explicit expressions
for the displacement field of an angular dislocation with one arm
perpendicular to the free surface and the other arm making an arbitrary
angle $\beta$ to the surface. Several typos are listed by Thomas
\citep{thomas93}. These expressions are simplified below for the
case of the second dislocation arm parallel to the surface, $\beta=\pi/2$.
In this case several indeterminate forms 0/0 have to be evaluated, so using
general formulae for $\beta=\pi/2$ would require some caution.

We use a coordinate system with the origin at the surface, the $X$-axis
parallel to the surface, the $Z$-axis along the inner surface normal,
the dislocation lies in the $XZ$-plane, the distance between the
surface and the dislocation arm parallel to it is $a$ {[}see Fig.\,\ref{fig:ConstructHalfloop}(a)
and Fig.\,1 in the main text of the paper{]}. A relation to three
coordinate systems introduced by Comninou and Dundurs \citep{Comninou1975}
is 
\begin{eqnarray}
X & = & y_{1}=\bar{y}_{1}=z_{3}=-\bar{z}_{3},\label{eq:0a}\\
Y & = & y_{2}=\bar{y}_{2}=z_{2}=\bar{z}_{2},\label{eq:0b}\\
Z & = & y_{3}+a=\bar{y}_{3}-a=a-z_{1}=\bar{z}_{1}-a.\label{eq:0c}
\end{eqnarray}

The following quantities are defined: 
\begin{equation}
R^{2}=X^{2}+Y^{2}+(Z-a)^{2},\,\,\,\,\,\bar{R}^{2}=X^{2}+Y^{2}+(Z+a)^{2},\label{eq:1}
\end{equation}

\begin{equation}
F=-\arctan\frac{Y}{X}-\arctan\frac{Y}{Z-a}-\arctan\frac{YR}{X(Z-a)},\label{eq:1a}
\end{equation}
\begin{equation}
\bar{F}=-\arctan\frac{Y}{X}+\arctan\frac{Y}{Z+a}+\arctan\frac{Y\bar{R}}{X(Z+a)}.\label{eq:4}
\end{equation}

Each component of the displacement vector is a sum of two terms, $u_{j}=u_{j}^{\infty}+u_{j}^{c}$
($j=1,2,3$). The first term is the displacement field of the dislocation
in the infinitely extended material plus its image, and the second
term releases the residual stress at the surface. The displacements
are written separately for each component of the Burgers vector $\mathbf{B}=(B_{1},B_{2},B_{3})$.
The $B_{2}$ and $B_{3}$ components of the Burgers vector correspond
to the edge and the screw threading dislocation arms involved in the
calculations of the present paper, and the corresponding formulae
for $B_{1}$ are included for the sake of completeness. 
\begin{widetext}
\textbf{Burgers vector (B$_{1}$,0,0)}

\begin{equation}
\frac{8\pi(1-\nu)}{B_{1}}u_{1}^{\infty}=2(1-\nu)(F+\bar{F})-XY\left\{ \frac{1}{R\left(R-(Z-a)\right)}+\frac{1}{\bar{R}\left(\bar{R}+Z+a\right)}\right\} \label{eq:5}
\end{equation}
\begin{equation}
\frac{8\pi(1-\nu)}{B_{1}}u_{2}^{\infty}=(1-2\nu)\left\{ \log\left(R-(Z-a)\right)+\log\left(\bar{R}+Z+a\right)\right\} -Y^{2}\left\{ \frac{1}{R\left(R-(Z-a)\right)}+\frac{1}{\bar{R}\left(\bar{R}+Z+a\right)}\right\} \label{eq:6}
\end{equation}

\begin{equation}
\frac{8\pi(1-\nu)}{B_{1}}u_{3}^{\infty}=Y\left\{ \frac{1}{R}-\frac{1}{\bar{R}}\right\} \label{eq:7}
\end{equation}

\begin{equation}
\frac{4\pi(1-\nu)}{B_{1}}u_{1}^{c}=-\frac{(1-2\nu)XY}{\left(\bar{R}+Z+a\right)^{2}}\left(\nu+\frac{a}{\bar{R}}\right)+\frac{XYZ}{\bar{R}\left(\bar{R}+Z+a\right)}\left\{ \frac{1}{\bar{R}+Z+a}\left(2\nu+\frac{a}{\bar{R}}\right)+\frac{a}{\bar{R}^{2}}\right\} \label{10}
\end{equation}

\begin{eqnarray}
\frac{4\pi(1-\nu)}{B_{1}}u_{2}^{c} & = & -\nu(1-2\nu)\log\left(\bar{R}+Z+a\right)-\frac{1-2\nu}{\bar{R}+Z+a}\left\{ \nu(Z+a)-a+\frac{Y^{2}}{\bar{R}+Z+a}\left(\nu+\frac{a}{\bar{R}}\right)\right\} \nonumber \\
 &  & +\frac{Z}{\bar{R}+Z+a}\left\{ -2\nu-\frac{a}{\bar{R}}+\frac{Y^{2}}{\bar{R}\left(\bar{R}+Z+a\right)}\left(2\nu+\frac{a}{\bar{R}}\right)+\frac{aY^{2}}{\bar{R}^{3}}\right\} \label{eq:11}
\end{eqnarray}

\begin{equation}
\frac{4\pi(1-\nu)}{B_{1}}u_{3}^{c}=\frac{2(1-\nu)Y}{\bar{R}+Z+a}\left(2\nu+\frac{a}{\bar{R}}\right)+\frac{YZ}{\bar{R}}\left(\frac{2\nu}{\bar{R}+Z+a}+\frac{a}{\bar{R}^{2}}\right)\label{eq:12}
\end{equation}

\textbf{Burgers vector (0,B$_{2}$,0)}

\begin{eqnarray}
\frac{8\pi(1-\nu)}{B_{2}}u_{1}^{\infty} & = & -(1-2\nu)\left\{ \log(R-(Z-a))+\log(\bar{R}+Z+a)\right\} +X^{2}\left\{ \frac{1}{R(R-(Z-a))}+\frac{1}{\bar{R}(\bar{R}+Z+a)}\right\} \nonumber \\
 &  & -\frac{Z-a}{R}+\frac{Z+a}{\bar{R}}\label{eq:14}
\end{eqnarray}

\begin{equation}
\frac{8\pi(1-\nu)}{B_{2}}u_{2}^{\infty}=2(1-\nu)(F+\bar{F})+XY\left\{ \frac{1}{R\left(R-(Z-a)\right)}+\frac{1}{\bar{R}\left(\bar{R}+Z+a\right)}\right\} -Y\left\{ -\frac{Z-a}{R\left(R-X\right)}+\frac{Z+a}{\bar{R}\left(\bar{R}-X\right)}\right\} \label{eq:15}
\end{equation}

\begin{equation}
\frac{8\pi(1-\nu)}{B_{2}}u_{3}^{\infty}=-(1-2\nu)\left\{ \log(R-X)-\log(\bar{R}-X)\right\} -X\left(\frac{1}{R}-\frac{1}{\bar{R}}\right)+\frac{(Z-a)^{2}}{R(R-X)}-\frac{(Z+a)^{2}}{\bar{R}(\bar{R}-X)}\label{eq:16}
\end{equation}

\begin{eqnarray}
\frac{4\pi(1-\nu)}{B_{2}}u_{1}^{c}	&=&	\nu(1-2\nu)\log\left(\bar{R}+Z+a\right)+\frac{1-2\nu}{\bar{R}+Z+a}\left\{ \nu(Z+a)-a+\frac{X^{2}}{\bar{R}+Z+a}\left(\nu+\frac{a}{\bar{R}}\right)\right\} \\
& &		+\frac{(1-2\nu)a}{\bar{R}}+\frac{Z}{\bar{R}+Z+a}\left\{ 2\nu+\frac{a}{\bar{R}}-\frac{X^{2}}{\bar{R}(\bar{R}+Z+a)}\left(2\nu+\frac{a}{\bar{R}}\right)-\frac{aX^{2}}{\bar{R}^{3}}\right\} +\frac{aZ(Z+a)}{\bar{R}^{2}(\bar{R}-X)}\left(\frac{X}{\bar{R}}-1\right) \label{eq:19}
\end{eqnarray}

\begin{eqnarray}
\frac{4\pi(1-\nu)}{B_{2}}u_{2}^{c}&=&\frac{(1-2\nu)XY}{\left(\bar{R}+Z+a\right)^{2}}\left(\nu+\frac{a}{\bar{R}}\right)-\frac{(1-2\nu)aY}{\bar{R}(\bar{R}-X)}+\frac{YZ}{\bar{R}\left(\bar{R}+Z+a\right)}\left\{ -\frac{2\nu X}{\bar{R}+Z+a}-\frac{aX}{\bar{R}}\left(\frac{1}{\bar{R}}+\frac{1}{\bar{R}+Z+a}\right)\right\} \\
& &+\frac{aYZ(Z+a)}{\bar{R}^{2}(\bar{R}-X)}\left(\frac{1}{\bar{R}-X}+\frac{1}{\bar{R}}\right)\label{eq:20}
\end{eqnarray}

\begin{eqnarray}
\frac{4\pi(1-\nu)}{B_{2}}u_{3}^{c} & = & -\frac{2(1-\nu)X}{\bar{R}+Z+a}\left(2\nu+\frac{a}{\bar{R}}\right)+\frac{2(1-\nu)a(Z+a)}{\bar{R}\left(\bar{R}-X\right)}+\frac{Z}{\bar{R}}\left\{ -\frac{2\nu X}{\bar{R}+Z+a}-\frac{aX}{\bar{R}^{2}}\right\} \nonumber \\
 &  & -\frac{aZ}{\bar{R}\left(\bar{R}-X\right)}\left[1-\frac{(Z+a)^{2}}{\bar{R}^{2}}-\frac{(Z+a)^{2}}{\bar{R}\left(\bar{R}-X\right)}\right]\label{eq:21}
\end{eqnarray}

\textbf{Burgers vector (0,0,B$_{3}$)} 
\begin{equation}
\frac{8\pi(1-\nu)}{B_{3}}u_{1}^{\infty}=Y\left\{ \frac{R-X}{R(R-X)}+\frac{\bar{R}-X}{\bar{R}(\bar{R}-X)}\right\} \label{eq:22}
\end{equation}
\begin{equation}
\frac{8\pi(1-\nu)}{B_{3}}u_{2}^{\infty}=(1-2\nu)\left\{ \log(R-X)+\log(\bar{R}-X\right\} -Y^{2}\left\{ \frac{1}{R(R-X)}+\frac{1}{\bar{R}(\bar{R}-X)}\right\} \label{eq:23}
\end{equation}
\begin{equation}
\frac{8\pi(1-\nu)}{B_{3}}u_{3}^{\infty}=2(1-\nu)(F-\bar{F})-Y\left\{ \frac{Z-a}{R(R-X)}+\frac{Z+a}{\bar{R}(\bar{R}-X)}\right\} \label{eq:24}
\end{equation}
\begin{equation}
\frac{4\pi(1-\nu)}{B_{3}}u_{1}^{c}=\frac{(1-2\nu)Y}{\bar{R}+Z+a}\left(1+\frac{a}{\bar{R}}\right)-\frac{YZ}{\bar{R}}\left(\frac{a}{\bar{R}^{2}}+\frac{1}{\bar{R}+Z+a}\right)\label{eq:27}
\end{equation}
\begin{eqnarray}
\frac{4\pi(1-\nu)}{B_{3}}u_{2}^{c} & = & (1-2\nu)\left\{ -\log(\bar{R}-X)-\frac{X}{\bar{R}+Z+a}\left(1+\frac{a}{\bar{R}}\right)+\frac{Z+a}{\bar{R}-X}\frac{a}{\bar{R}}\right\} +\frac{XZ}{\bar{R}}\left(\frac{a}{\bar{R}^{2}}+\frac{1}{\bar{R}+Z+a}\right)\nonumber \\
 &  & -\frac{Z}{\bar{R}-X}\left\{ -\frac{a}{\bar{R}}+\frac{Z+a}{\bar{R}}\left(1+\frac{a(Z+a)}{\bar{R^{2}}}\right)+\frac{a(Z+a)^{2}}{\bar{R}^{2}(\bar{R}-X)}\right\} \label{eq:28}
\end{eqnarray}
\begin{equation}
\frac{4\pi(1-\nu)}{B_{3}}u_{3}^{c}=2(1-\nu)\left\{ \bar{F}+\frac{Y}{\bar{R}-X}\frac{a}{\bar{R}}\right\} +\frac{YZ}{\bar{R}(\bar{R}-X)}\left\{ 1+\frac{Z+a}{\bar{R}-X}\frac{a}{\bar{R}}+\frac{a(Z+a)}{\bar{R}^{2}}\right\} \label{eq:29}
\end{equation}
\end{widetext}

\section{Displacement field of a straight dislocation perpendicular to the
surface}

\label{sec:ThreadingDisl}

We present here expressions for the displacement fields of the edge
and screw straight dislocations perpendicular to the surface of an
elastic half-space, see Eqs.\,(213) and (220) in Ref.\,\citep{lothe92}.
The sums of these displacement fields and those above for the angular
dislocations with the Burgers vectors $(0,B_{2},0)$ and $(0,0,B_{3})$
give the L-shaped dislocation in Fig.~\ref{fig:ConstructHalfloop}(a).
To do this, the direction of the $Z$-axis is reversed with respect
to Ref.\,\citep{lothe92}, and the displacement field for the edge
dislocation is rotated by $90^{\circ}$. We define
\begin{equation}
r^{2}=X^{2}+Y^{2}+Z^{2}.\label{eq:30}
\end{equation}
Then,
\begin{widetext}
\textbf{Burgers vector (0,B$_{2}$,0)
\begin{equation}
\frac{4\pi(1-\nu)}{B_{2}}u_{1}=(1-2\nu)\log\sqrt{X^{2}+Y^{2}}+\frac{Y^{2}}{X^{2}+Y^{2}}-\nu\left[(1-2\nu)\log(r+Z)+(3-2\nu)\left(\frac{Z}{r+Z}+\frac{X^{2}}{(r+Z)^{2}}\right)-\frac{2X^{2}}{r(r+Z)}\right]\label{eq:31}
\end{equation}
\begin{equation}
\frac{4\pi(1-\nu)}{B_{2}}u_{2}=2(1-\nu)\arctan\frac{Y}{X}-\frac{XY}{X^{2}+Y^{2}}-\nu\left[(1-2\nu)\frac{XY}{(r+Z)^{2}}-\frac{2XYZ}{r(r+Z)^{2}}\right]\label{eq:32}
\end{equation}
}

\begin{equation}
\frac{2\pi(1-\nu)}{B_{2}}u_{3}=\nu X\left[\frac{1}{r}+(1-2\nu)\frac{1}{r+Z}\right]\label{eq:33}
\end{equation}

\textbf{Burgers vector (0,0,B$_{3}$)} 
\begin{equation}
\frac{2\pi}{B_{3}}u_{1}=-\frac{Y}{r+Z},\,\,\,\,\,\frac{2\pi}{B_{3}}u_{2}=\frac{X}{r+Z},\,\,\,\,\,\frac{2\pi}{B_{3}}u_{3}=\arctan\frac{Y}{X}\label{eq:34}
\end{equation}
\end{widetext}

\section{X-ray diffraction intensity in the Stokes-Wilson approximation}

Stokes and Wilson \citep{StokesWilson44} showed that the X-ray diffraction
intensity distribution in a highly distorted crystal is equal to the
probability density distribution of the strain. The limits of applicability
of this approximation were considered in Ref.\,\citep{kaganer14acta}.
It was shown that the Stokes-Wilson approximation is applicable as
long as the long range order is not seen as a coherent peak in the
diffraction profiles. The relevant strain components depend on the
diffraction geometry. Stokes and Wilson developed their approximation
for powder diffraction, in which case the normal strain in the direction
of the diffraction vector is involved. The corresponding components
for the reciprocal space maps and skew diffraction geometry of singe
crystals are derived below.

\subsection{Reciprocal space maps}

A general expression for the X-ray diffraction intensity from a crystal
containing lattice defects is
\begin{equation}
I(\mathbf{q})=\iint G(\mathbf{r}_{1},\mathbf{r}_{2})\exp\left[i\mathbf{q}\cdot(\mathbf{r}_{2}-\mathbf{r}_{1})\right]d\mathbf{r}_{1}d\mathbf{r}_{2},\label{eq:35}
\end{equation}
where the integration is performed over the volume of the crystal,
$\mathbf{r}_{1}$ and $\mathbf{r}_{2}$ are two points inside the
crystal and $\mathbf{q}$ is a small deviation of the diffraction
vector $\mathbf{Q}$ from the reciprocal lattice point. The correlation
function $G(\mathbf{r}_{1},\mathbf{r}_{2})$ is
\begin{equation}
G(\mathbf{r}_{1},\mathbf{r}_{2})=\left\langle \exp\left\{ i\mathbf{Q}\cdot\left[\mathbf{U}(\mathbf{r}_{2})-\mathbf{U}(\mathbf{r}_{1})\right]\right\} \right\rangle ,\label{eq:36}
\end{equation}
where $\mathbf{U}(\mathbf{r}_{1})$ and $\mathbf{U}(\mathbf{r}_{2})$
are total displacements at these points due to all defects in the
crystal and $\left\langle \ldots\right\rangle $ denotes a statistical
average over the distribution of the defects.

For an epitaxial film with the $z$-axis along the film normal, the
correlation function can be written as $G(x_{2}-x_{1},y_{2}-y_{1},z_{1},z_{2})$,
due to a translational invariance in the lateral plane (but not in
the $z$-direction). We consider reciprocal space maps measured in
a standard triple-crystal laboratory X-ray diffractometer. The incident
and diffracted X-rays are collimated in the scattering plane $(q_{x},q_{z})$
and integrated over the wave vector component $q_{y}$ normal to it
(the ``vertical divergence'' in a standard diffractometer setup with
horizontal scattering plane). The integration of the $q_{y}$-dependent
term $\exp\left[iq_{y}\left(y_{2}-y_{1}\right)\right]$ in Eq.\,(\ref{eq:35})
over $q_{y}$ in infinite limits gives a delta function $\delta\left(y_{2}-y_{1}\right)$,
and the integral (\ref{eq:35}) is simplified to
\begin{eqnarray}
I(q_{x},q_{z}) & = & \intop_{-\infty}^{\infty}dx\iintop_{0}^{t}dz_{1}\,dz_{2}\,G\left(x,z_{1},z_{2}\right)\nonumber \\
 &  & \times\exp\left[iq_{x}x+iq_{z}\left(z_{2}-z_{1}\right)\right].\label{eq:37}
\end{eqnarray}
The argument $y=0$ in the correlation function is omitted hereafter
for simplicity.

For high dislocation density, only correlations between closely
spaced points are important. The difference of displacements in Eq.\,(\ref{eq:36})
is then approximated as $\mathbf{Q}\cdot\left[\mathbf{U}(\mathbf{r}_{2})-\mathbf{U}(\mathbf{r}_{1})\right]\approx\kappa_{x}x+\kappa_{z}\varsigma$,
where $\varsigma=z_{2}-z_{1}$, $\kappa_{x}=\partial(\mathbf{Q}\cdot\mathbf{U})/\partial x$,
and $\kappa_{z}=\partial(\mathbf{Q}\cdot\mathbf{U})/\partial z$.
Then, the statistical mean (\ref{eq:36}) can be written as
\begin{equation}
G(x,\varsigma,z)=\iintop_{-\infty}^{\infty}P(\kappa_{x},\kappa_{z},z)\exp\left(i\kappa_{x}x+i\kappa_{z}\varsigma\right)d\kappa_{x}\,d\kappa_{z}.\label{eq:38}
\end{equation}
Here $P(\kappa_{x},\kappa_{z},z)$ is the joint probability distribution
of the respective strain components taken at a depth $z$. The integral
over $z_{1}$ and $z_{2}$ in Eq.\,(\ref{eq:37}) can be written
as an integral over $z$ and $\varsigma$, and the latter integral
can be extended in the infinite limits, since only small $\varsigma$
are relevant. Then the Fourier integral (\ref{eq:37}) gives 
\begin{equation}
I(q_{x},q_{z})=\intop_{0}^{t}P\left[q_{x}=-\frac{\partial(\mathbf{Q}\cdot\mathbf{U})}{\partial x},q_{z}=-\frac{\partial(\mathbf{Q}\cdot\mathbf{U})}{\partial z},z\right]dz.\label{eq:39}
\end{equation}
This equation replaces Eq.\,(10) in Ref.\,\citep{kaganer14acta}
where the product of probabilities is written instead of the joint
probability.

The Monte Carlo implementation of Eq.\,(\ref{eq:39}) is straightforward.
First, a pixel array is defined for $I(q_{x},q_{z})$ to cover the
range of the wave vectors of interest. Then dislocations are generated
according to their density and distribution. The sum of their displacements
is used to calculate the strain at the point $(0,0,z)$, where $z$
is uniformly seeded from $0$ to $t$. Since the analytical differentiation
of the displacements for an angular dislocation presented above would
lead to very bulky expressions, we calculate derivatives of the displacements
required in Eq.\,(\ref{eq:39}) from a difference of the displacements
at two closely spaced points. After calculating the strain components
$q_{x}$ and $q_{z}$ in Eq.\,(\ref{eq:39}), 1 is added to the corresponding
pixel of the array $I(q_{x},q_{z})$. The dislocation generation is
repeated.

\subsection{Skew diffraction geometry}

\begin{figure*}
\includegraphics[width=1\textwidth]{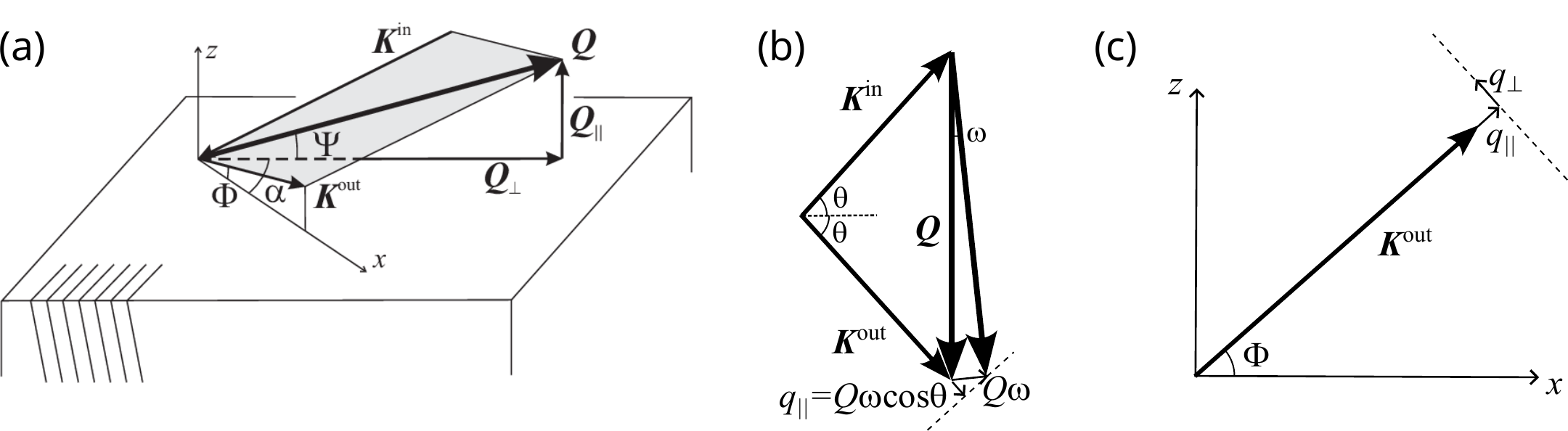}

\caption{Sketch of skew diffraction geometry: (a) three-dimensional picture
reproduced from Ref.\,\citep{kaganer05GaN}, (b) wave vectors in
the scattering plane, (c) wave vectors in the plane normal to the
surface and containing the scattered beam direction $\mathbf{K}^{\mathrm{out}}.$}

\label{fig:SkewSketch}
\end{figure*}

Figure \ref{fig:SkewSketch}(a) reproduces a sketch of the skew diffraction
geometry from Ref.\,\citep{kaganer05GaN}. The details of the geometry
and the definition of the angles can be found in the cited paper.
Our aim now is to average the intensity (\ref{eq:35}) over the plane
perpendicular to the direction of the wave vector of the scattered
wave $\mathbf{K}^{\mathrm{out}}$. Figure \ref{fig:SkewSketch}(b)
shows the scattering plane (the plane containing the wave vectors
of the incident $\mathbf{K}^{\mathrm{in}}$ and the diffracted $\mathbf{K}^{\mathrm{out}}$
waves). On sample rotation by an angle $\omega$, the wave vector
$\mathbf{q}$ is directed perpendicular to $\mathbf{Q}$ and has a
length $q=Q\omega$. Its component along the diffracted beam direction
is $q_{\parallel}=Q\omega\cos\theta$, where $\theta$ is the Bragg
angle.

The coordinates in Fig.\,\ref{fig:SkewSketch}(a) are chosen such that
the $xz$ plane is perpendicular to the surface and contains the wave
vector $\mathbf{K}^{\mathrm{out}}$. The wave vectors in this plane
are shown in Fig.\,\ref{fig:SkewSketch}(c). The integration of the
intensity over the plane perpendicular to $\mathbf{K}^{\mathrm{out}}$
is the integration over $q_{y}$ perpendicular to this plane and the
integration over $q_{\perp}$ in this plane. The first integration
is carried out as above and gives Eq.\,(\ref{eq:37}) with the present
choice of the axis directions. To perform the second integration,
we express the wave vectors as
\begin{eqnarray}
q_{x} & = & q_{\parallel}\cos\Phi-q_{\perp}\sin\Phi,\label{eq:40}\\
q_{z} & = & q_{\parallel}\sin\Phi+q_{\perp}\cos\Phi,\nonumber 
\end{eqnarray}
so that
\begin{eqnarray}
q_{x}x+q_{z}(z_{2}-z_{1}) & = & \left[x\cos\Phi+(z_{2}-z_{1})\sin\Phi\right]q_{\parallel}\label{eq:41}\\
 & + & \left[-x\sin\Phi+(z_{2}-z_{1})\cos\Phi\right]q_{\perp}.\nonumber 
\end{eqnarray}
Integrating the intensity (\ref{eq:37}) over $q_{\perp}$ gives rise
to a delta function $\delta\left(-x\sin\Phi+(z_{2}-z_{1})\cos\Phi\right)$.
Substituting $x=(z_{2}-z_{1})\cot\Phi$ in Eq.\,(\ref{eq:41}) gives
\begin{equation}
q_{x}x+q_{z}(z_{2}-z_{1})=q_{\parallel}(z_{2}-z_{1})/\sin\Phi,\label{eq:42}
\end{equation}
and the diffracted intensity is
\begin{eqnarray}
I(q_{\parallel}) & = & \iintop_{0}^{t}dz_{1}\,dz_{2}\,G\left((z_{2}-z_{1})\cot\Phi,z_{1},z_{2}\right)\nonumber \\
 &  & \times\exp\left[iq_{\parallel}(z_{2}-z_{1})/\sin\Phi\right].\label{eq:43}
\end{eqnarray}
This equation coincides with Eq.\,(6) in Ref.~\citep{kopp13jac}.

In the limit of threading dislocations in an infinitely thick crystal,
the correlation function in Eq\@.\,(\ref{eq:43}) can be written
as $G\left((z_{2}-z_{1})\cot\Phi\right)$, since the displacements
become $z$-independent. Substituting $\xi=z\cot\Phi$ reduces this
equation to
\begin{equation}
I(q_{\parallel})=\intop_{-\infty}^{\infty}G(\xi)\exp\left(iq_{\parallel}\xi/\cos\Phi\right)d\xi.\label{eq:44}
\end{equation}
This equation coincides with Eq.~(9) in Ref.\,\citep{kaganer05GaN}.

In the Stokes-Wilson approximation for Eq.\,(\ref{eq:43}), we write
\begin{eqnarray}
\mathbf{Q}\cdot\left[\mathbf{U}(\mathbf{r}_{2})-\mathbf{U}(\mathbf{r}_{1})\right] & \approx & \frac{\partial(\mathbf{Q}\cdot\mathbf{U})}{\partial x}(z_{2}-z_{1})\cot\Phi\nonumber \\
 & + & \frac{\partial(\mathbf{Q}\cdot\mathbf{U})}{\partial z}(z_{2}-z_{1})\\
 & = & \kappa\varsigma/\sin\Phi,\label{eq:45}
\end{eqnarray}
where again $\varsigma=z_{2}-z_{1}$ and
\begin{equation}
\kappa=\cos\Phi\frac{\partial(\mathbf{Q}\cdot\mathbf{U})}{\partial x}+\sin\Phi\frac{\partial(\mathbf{Q}\cdot\mathbf{U})}{\partial z}.\label{eq:46}
\end{equation}
Since the wave vector of the diffracted beam $\mathbf{K}^{\mathrm{out}}$
makes an angle $\Phi$ to the $x$ axis, this equation can be written
as
\begin{equation}
\kappa=\mathbf{\hat{K}}^{\mathrm{out}}\cdot\nabla\left(\mathbf{Q}\cdot\mathbf{U}\right),\label{eq:47}
\end{equation}
where $\mathbf{\hat{K}}^{\mathrm{out}}$ is the unit vector in the
direction of $\mathbf{K}^{\mathrm{out}}$. The statistical average
in the correlation function (\ref{eq:36}) can be written as
\begin{equation}
G(\varsigma,z)=\intop_{-\infty}^{\infty}P(\kappa,z)\exp(\kappa\varsigma/\sin\Phi)\,d\kappa.\label{eq:48}
\end{equation}
The integral (\ref{eq:43}) then gives
\begin{equation}
I(q_{\parallel})=\intop_{0}^{t}P\left(q_{\parallel}=-\mathbf{\hat{K}}^{\mathrm{out}}\cdot\nabla\left(\mathbf{Q}\cdot\mathbf{U}\right),z\right)dz.\label{eq:49}
\end{equation}
For infinitely long threading dislocations, when distortions do not
depend on $z$, this equation reduces to Eq.\,(2) in Ref.\,\citep{kaganer15jpd}.

\begin{figure}
\includegraphics[width=1\columnwidth]{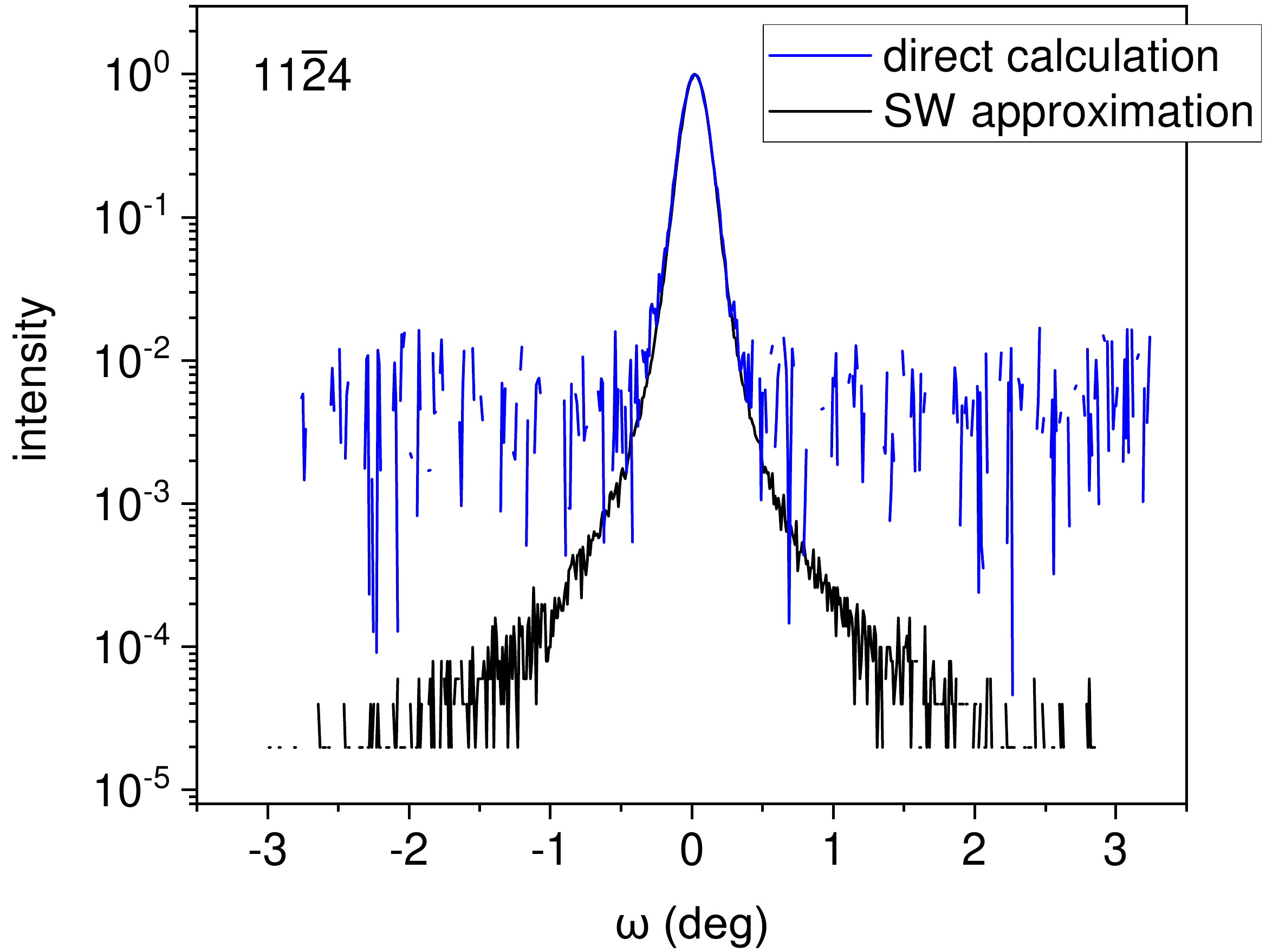}

\caption{X-ray diffraction profile in $11\bar{2}4$ reflection in skew geometry
for dislocation half-loops with edge threading arms. Threading dislocation
density $\rho_{\mathrm{T}}=1\times10^{10}$cm$^{-2}$, mean length
of the misfit segments $L=1$ \textmu m, film thickness $t=1$ \textmu m.
The blue curve is a calculation using the dislocation displacements
given by Eqs.\,(\ref{eq:35}), (\ref{eq:36}), while the black curve
is a calculation in the Stokes-Wilson approximation (\ref{eq:49})
using the dislocation strain. The curves are calculated by averaging
over $N=2\times10^{6}$ and $1\times10^{6}$ random dislocation arrangements,
respectively, which took the same CPU time.}

\label{fig:EversusU}
\end{figure}

Figure \ref{fig:EversusU} compares double crystal diffraction profiles
in skew geometry in GaN($11\bar{2}4$) reflection calculated by the
Monte Carlo method using Eqs.\,(\ref{eq:35}) and (\ref{eq:36})
with the strain probability density calculation in the Stokes-Wilson
approximation (\ref{eq:49}). The calculation of displacements for
angular dislocations by the formulae presented in Sec.\,\ref{sec:AngularDisl}
takes the largest part of the calculation time. Since the strain is
calculated from the difference of the displacements in two close points,
we compare the average over $2\times10^{6}$ and $1\times10^{6}$
random dislocation arrangements for the displacement-based and the
strain-based calculations respectively, since they require the same
CPU time. Each of the calculations shown in Fig\@.\,\ref{fig:EversusU}
took 20~minutes CPU time when running on 128 cores.

The accuracy achieved in the calculations is evident from the noise
levels of the respective curves: the accuracy achieved in the strain-based
calculation is at least a hundred times higher. Since the accuracy improves
as $1/\sqrt{N}$, where $N$ is the number of repetitions, it would
take at least $10^{4}$ more computing time to achieve the same accuracy
with the displacement based calculation.


%